\documentclass[12pt]{article}

\usepackage{indentfirst}
\usepackage{amsmath}
\usepackage{amssymb}
\usepackage{amsfonts}
\usepackage{amscd}
\usepackage{amsbsy}
\usepackage{amsthm}
\usepackage{latexsym}
\usepackage{graphicx,color} 
\usepackage[dvipsnames]{xcolor}
\usepackage[colorlinks]{hyperref}
\hypersetup{linkcolor=blue,citecolor=blue,urlcolor=blue}
\def\nn{\nonumber}       
\def\beq{\begin{eqnarray}}
\def\eeq{\end{eqnarray}}

\def\diag{\,\mbox{diag}\,}

\def\al{\alpha}
\def\be{\beta}
\def\ch{\chi}
\def\ga{\gamma}
\def\de{\delta}
\def\vp{\varepsilon}
\def\ep{\epsilon}

\def\la{\lambda}
\def\na{\nabla}
\def\pa{\partial}

\def\si{\sigma}
\def\om{\omega}
\def\ph{\varphi}

\def\th{\theta}

\def\up{\upsilon}
\def\Ga{\Gamma}

\def\Si{\Sigma}

\usepackage[symbol]{footmisc} 
\usepackage{float} 
\usepackage{cite}  
\usepackage{titlesec}

\usepackage{ulem} 

\titleformat*{\section}{\large\bfseries}
\titleformat*{\subsection}{\normalsize\bfseries}

\begin{document}
	
	\begin{center}
		
{\Large
Pauli equation and charged spin-$1/2$ 
particle
\\
in a weak gravitational field}
\vskip 6mm
		
		\textbf{Samuel W. P. Oliveira}
		\footnote{E-mail address: sw.oliveira55@gmail.com}$^{a,b}$,
		\quad
		\textbf{Guilherme Y. Oyadomari}
		\footnote{E-mail address: yoshioyadomari@hotmail.com}$^{a,b}$,
		\quad
		\textbf{Ilya L. Shapiro}\footnote{
E-mail address: ilyashapiro2003@ufjf.br}$^{b,a}$
		\vskip 6mm
		
$a.$ \
PPGCosmo,
		Universidade Federal do Esp\'{\i}rito Santo
\\
		Vit\'oria, 29075-910, ES, Brazil
\vskip 3mm
		
$b.$ \ 
		Departamento de F\'{\i}sica, \ ICE, \
		Universidade Federal de Juiz de Fora
		\\ Juiz de Fora,  36036-900,  MG,  Brazil
	\end{center}
	\vskip 5mm
	
	\centerline{\textbf{Abstract}}
	
	\begin{quotation}
		
\noindent
Using the nonrelativistic approximation in the curved-space Dirac
equation, the analog of the Pauli equation is derived for a weak
gravitational field with a gauge fixing condition related to the 
synchronous gauge, in the presence of an
electromagnetic field. Different from the previous works which
were employing either the exact or conventional Foldy-Wouthuysen
transformations, here we perform calculations by directly performing
nonrelativistic approximation which reduced in the power series
expansion in the inverse mass of the spinning particle. On top of
that, the equations of motion for the massive spin-$1/2$ charged
particle are obtained.
   The two particular cases of the previously explored backgrounds,
   namely a) plane gravitational wave and b) homogeneous static
   gravitational field are considered for control. In the case a) we
   meet correspondence with the previous results. On the other hand,
   in case b), there is no correspondence with neither perturbative nor
   with exact Foldy-Wouthuysen	transformations, which we also
   recalculate and agree with the previous works. The disagreement
   is a kind of a theoretical challenge and most likely occurs because
the potential energy, in the particular case of Newtonian
approximation, is proportional to the mass of the particle.
\vskip 2mm

\vskip 2mm
		
	\end{quotation}


	\newpage
	
	\section{Introduction}
	\label{s1}

	General relativity (GR) is a generalization of Newton's laws of
	classical mechanics, in particular it makes classical law of gravity
	compatible with the relativistic mechanics at low energies (in the
	IR). The main symmetry of GR is the general covariance and, in
	particular, the construction of fields in curved spacetime of GR
	can be achieved on the basis of this symmetry. E.g., this is
	the case for the Dirac field describing the spin-1/2 particle. This
	means, the construction of  the action of Dirac field in curved
	background is essentially based on covariance (see, e.g.,
	\cite{DeWitt2003,Rubakov,ParToms,MensAgitat,OUP}).
	Thus, the nonrelativistic approximation,
	in particular the Pauli equation for the spin-1/2 particle, in
	curved space can be constructed only starting from the covariant
	Dirac equation. When making a nonrelativistic approximation,
	the covariance is lost, making direct generalization of the Pauli
	equation to the curved spacetime impossible.
	
	The nonrelativistic approximation has great importance in
	relativistic quantum mechanics and quantum field theory (QFT). The
	main reason is that, from the usual QFT perspective, many relevant
	physical applications correspond to the low-energy domain, where one
	can meet many high-precision experiments. This statement is valid
	also for the gravitational experiments, in the sense that the highest
	precision measurements are done in laboratory and for the motions
	with the nonrelativistic speeds. Clear examples of this can be found
	in the kinds of existing and forthcoming experiments  (see, e.g.,
	\cite{Long-2003,South}).   From the side of the theory, it looks
	necessary to have well established gravitational contributions to the
	Pauli equations and to its classical counterpart, i.e., the equations
	describing the nonrelativistic spin-1/2 charged particle.
	
	Many of the existing literature on relativistic generalizations of
	the nonrelativistic Schrodinger equation for spinning particle
	with gravity discusses the gravitational field created by the
	accelerated reference frames, the constant gravitational
	acceleration $\vec{g}$, usually obtained by the corresponding
	approximation in the Schwarzschild metric \cite{Obukhov-2000}
	(see also earlier works \cite{Bonse-1983,Mashhoon,Silenko-2005}), further
	generalizations, including the  rotations
	\cite{Silenko-2006,Obukhov-2009,Gosselin-2010} and the
	motion of spin in different configurations of the gravitational field
	\cite{Obukhov-2013} (see also \cite{Obukhov-2017,Quach-2020}).
	Another type of gravitational field
	considered in the literature is the one of a flat gravitational wave
	\cite{Goncalves-2007}. This paper found that this kind of the
	gravitational field admits an exact Foldy-Wouthuysen transformation
	(there were also further developments in \cite{Goncalves-2018}).
	However, soon after it was noted that the corresponding calculation
	had a mistake leading to the discrepancy with the conventional
	Foldy-Wouthuysen transformation \cite{Quach-2015}.
	
	All of the mentioned works were done using either usual
	perturbative or exact Foldy-Wouthuysen transformation. Since the
	equations in different papers sometimes differ, it makes sense to
	verify the results by a qualitatively different and simpler method,
	such as the simplest textbook-level one-step derivation of the Pauli
	equation \cite{LL-4,BjorkenDrell-1}. In the rest of this paper we
	shall present, in many details, of such a simple calculation on an
	arbitrary weak gravitational field. The main restrictions on the
	background are that the metric can be presented in the form
	\beq
	g_{\mu\nu}\,=\,\eta_{\mu\nu} + h_{\mu\nu},
	\label{hmn}
	\eeq
	leaving aside the rotational metrics. As it was mentioned above, we
	assume the weak field restriction, that means $|h_{\mu\nu}| \ll 1$
	and the same for its partial derivatives.
	
	In what follows, we obtain the gravitational version of the Pauli
	equation for the metric (\ref{hmn}) and an arbitrary electromagnetic
	field $\big({\vec E},{\vec B}\big)$. Furthermore, the canonical
	quantization of this equation and the subsequent classical limit
	gives the equations for a spinning particle. These two main results
	are applied to the particular background of a plane gravitational
	wave and Newtonian gavity, which were worked out in
	\cite{Silenko-2005} and  \cite{Goncalves-2007,Quach-2015}.
	This comparison provides a good portion of control and corrections
	of coefficients in some cases.
	
	The paper is organized as follows. In Sec.~\ref{s2} we briefly review
	the construction of the covariant derivative of a Dirac fermion and
	of the corresponding action and give necessary expressions for its
	expansion in the weak gravity limit.  Sec.~\ref{s3} presents the
	detailed derivation of the Pauli equation.  Sec.~\ref{s4} reports on
	the calculation of the equations of motion on a background of weak,
	but otherwise almost arbitrary, gravitational and electromagnetic,
	fields. The restriction is related to the synchronous gauge fixing
	condition. Different from cosmology, here this condition makes
	the results less general. On the other hand, it greatly reduce the
	amount of the (already significant) calculations.
	In Sec.~\ref{s5}, there is a comparison of our results with
	the results of \cite{Goncalves-2007,Quach-2015} concerning the
	gravitational waves background. Sec.~\ref{s6} makes similar
	comparison with the Newtonian limit of the Schwarzschild metric
	\cite{Silenko-2005}. Finally, the Sec.~\ref{ConcDisc}, we draw our
	conclusions.

	\section{Dirac equation in a gravitational field background}
	\label{s2}
	
	Consider briefly the formulation of the action of Dirac fermion.
	More details can be found  in \cite{MensAgitat}. The consideration
	which is based on a similar approach can be found, 	e.g., in
\cite{ParToms,MensAgitat} and the treatment based on group
	theory, e.g., in \cite{OUP}. The form of the action is
	\beq
	S_f
	&=&
	\frac{i}{2}\int d^{4}x \sqrt{-g}\big(
	\bar{\psi} \ga^{\mu}\na_{\mu}\psi
	- \na_{\mu}\bar{\psi}\,\ga^\mu\psi + 2im \big)\,.
	\label{25}
	\eeq
	To make this definition more concrete, we need to construct a
	covariant version of the gamma-matrices $\ga^{\mu}$ and the
	covariant derivative of a fermion $\na_\mu$. We start by
	introducing the tetrads $e^\mu_a$ and $e_\nu^b$,
	\beq
	\label{1}
	e^{\mu} _a e^{a\nu}
	= e^\mu _a e^\nu _b \eta^{ab} = g ^{\mu \nu},
	\qquad
	e^{a} _{\mu} e_{a\nu}
	= g_{\mu \nu},
	\quad
	e^a _\mu e^\mu _b
	= \de^a _b ,
	\quad
	e^a_\mu e^\al_{a} = \delta^\al _\mu .
	\eeq
	In these and subsequent formulas, Greek letters, e.g.,
	$\mu=0,1,2,3$ are covariant indices corresponding to the
	coordinates of the Riemannian space. Latin letters, such
	as $a=0,1,2,3$, are Lorentz indices corresponding to the coordinates
	of the flat Minkowski space which is a tangent space to the manifold
	in the given point. For an arbitrary vector $V^\mu$, the
	transformation formulas are $V^\mu=e^{\mu} _a V^a$ and
	$V^b=e_\nu^b V^\nu$.
	
	For the gamma matrices we define $\,\ga^\mu = e^\mu _a \ga^a$.
As usual, $\ga^5=i\ga^0\ga^1\ga^2\ga^3$.
Under the diffeomorphism
	transformations, $\,\ga^\mu\,$ and $\,\ga_\nu\,$ are contravariant
	and covariant components of a vector. These objects satisfy the
	covariant version of Clifford algebra,
	\beq
	\ga_\mu\ga_\nu + \ga_\nu\ga_\mu &=& g_{\mu\nu}.
	\label{Cliff}
	\eeq
	
	The covariant derivative of a spinor is defined by the relations
	\beq
	\na_\mu \psi
	&=&
	\pa_\mu \psi
	+ \frac{i}{2}\,
	\om^{\,\,\,ab}_{\mu\, \cdot \cdot}\,\si_{ab}\,\psi \,,
	\quad
	\mbox{where}
	\quad
	\si_{ab} = \frac{i}{2} \big(\ga_a \ga_b - \ga_b \ga_a\big)\,.
	\label{co1}
	\eeq
	The complex conjugation gives
	\beq
	\na_\mu \bar{\psi}
	&=&
	\pa_\mu  \bar{\psi} - \frac{i}{2}\,
	\om^{\,\,\,ab}_{\mu\, \cdot \cdot}\,  \bar{\psi} \si_{ab}\,.
	\label{co1conj}
	\eeq
	In this expressions, $\,\om^{\,\,\,ab}_{\mu\, \cdot \cdot}\,$  are the
	real-valued coefficients of the {\it spinor connection}, which have
	to be found from the requirement of covariance.
	
	The solution for the spinor connection is
	\beq
	\om^{\,\,\,ab}_{\mu\, \cdot \cdot}
	&=&
	\frac{1}{2}\,\big(e^{b}_\tau e^{\la a}
	\Ga^\tau_{\la \mu} - e^{\la a}\pa_{\mu}e^{b}_{\la}\big)
	\,=\,-\,\om^{\,\,\,ba}_{\mu\, \cdot \cdot} \,\,.
	\label{spishort}
	\eeq
	It is sometimes useful to use the explicitly
	antisymmetric form,
	\beq
	\om^{\,\,\,ab}_{\mu\, \cdot \cdot}
	&=&
	\frac14\,\big(e^{b}_\tau e^{\la a} - e^{a}_\tau e^{\la b}\big)
	\Ga^\tau_{\la \mu}
	\,+\,
	\frac{1}{4}\big(e^{\la b}\pa_{\mu}e_{\la}^{a}
	- e^{\la a}\pa_{\mu}e^{b}_{\la}\big)\,.
	\label{spi}
	\eeq
	
	We need to consider the expansion of the action (\ref{25})
	on the flat background (\ref{hmn}). For the sake of generality,
	we can work out a more general expansion
	\beq
	g_{\al\be} \, \, \,\longrightarrow  \, \, \, g^\prime_{\al\be}
	\,= \,
	g_{\al\be} + h_{\al\be}.
	\label{var}
	\eeq
	The first orders of expansions can be easily found in the form
	\beq
	\nonumber
	\de \sqrt{- g} &=& \frac{1}{2}\sqrt{- g} h\,,
	\quad
	\de g^{\mu\nu} = - h^{\mu\nu}\,,
	\quad
	\de e^{c}_\mu = \frac12\,h^\nu_\mu e^c_\nu\,,
	\quad
	\de e^\rho_b = - \frac12\,h^{\rho}_{\la} e^\la_{b}\,,
	\\
	\label{26}
	\de \Ga^{\la}_{\al \be}
	&=&
	\frac12\,
	\big(
	\na_{\al}h^\la_\be + \na_\be h^\la_\al - \na^\la
	h_{\al \be}\big)\,,
	\quad
	\de \ga^\mu = - \frac12\,h^\mu_\nu\,\ga^\nu
	\eeq
	and
	\beq
	\de \om^{\,\,\,ab}_{\mu \,\cdot \cdot}
	\,=\,
	\frac12\,
	\big(e^{a\tau}e^{b\la} - e^{b\tau}e^{a\la}\,)
	\na_{\la}h_{\mu \tau}
	\label{27}
	\eeq
	Here all indices are raised and lowered with the background
	metrics $g^{\mu\nu}$ and $g_{\mu\nu}$ and the covariant
	derivatives are constructed with the background metric
	$g_{\mu\nu}$ and the corresponding connection.
	It can be shown that the variation of the spinor
	connection (\ref{27}) is irrelevant, as its contribution vanishes
	when substituted into the variation of (\ref{25}).
	
	Using Eqs.~(\ref{26}), we arrive at the first-order variation of
	the action (\ref{25})
	\beq
	\de S_f
	&=&
	\int d^4x \sqrt{-g}\,
	h_{\al \be}\,
	\Big\{ \frac{i}{4} g^{\al \be}
	\big(
	\bar{\psi}\ga^\la \na_\la \psi - \na_\la\bar{\psi} \ga^\la\psi
	\big)
	\nonumber
	\\
	&&
	-\,\,\frac{i}{4}\,\big(\bar{\psi}\ga^\al\na^\be\psi
    - \na^\al \bar{\psi}\,\ga^\be \psi\big)
	- \frac12\,g^{\al \be} m \,\bar{\psi}\psi
	\Big\}\,,
	\label{actionexp}
	\eeq
	where $h=h^\mu_{\,\mu}$. Taking the flat background metric,
	the last expression is a good starting point to perform the
	nonrelativistic expansion in a weak gravitational field.
	
	\section{Pauli equation in a weak gravitational field}
\label{s3}	

The standard approach to the nonrelativistic approximation is that
it assumes weak external fields because a strong field can accelerate
a particle to become relativistic. Indeed, the weak gravitational
field condition does not reduce to requiring $|h_{\mu\nu}| \ll 1$
but, as we already mentioned above, demands the same from the
partial derivatives of $h_{\mu\nu}$. Following this logic, we come
back to the expansion of the metric on a flat background (\ref{hmn})
and arrive at the simplified version of (\ref{actionexp}), which is
the starting action of the nonrelativistic approximation
\beq
S &=& \int d^4x
\Big\{
\bar{\psi} \left( i \ga^{\al} \pa_{\al} - m \right) \psi
\nn
\\
&&
+ \,\,\dfrac{i}{4} \left(  h g^{\al \be} - h^{\al \be} \right)
\left( \bar{\psi}\, \ga_{\be}\, \pa_{\al} \psi - \pa_{\al}
\bar{\psi}\, \ga_{\be}\, \psi \right)
\,-\,\dfrac{mh}{2} \bar{\psi} \psi \Big\},
\label{eq1}
\eeq
where the first term is the basic zeroth-order part.
Starting from this point the background metric is $\eta^{\al \be}$,
but we use Greek letters for the relativistic indices, e.g.,
$\al = 0,\,1,\,2,\,3$, while the Latin indices are space $3D$, 
e.g., $i,j,k,.. = 1,\,2,\,3$. The signature is
$\,\eta_{\mu \nu} = \diag (1,-1,-1,-1) $.

In the given approximation, the Dirac equation reads
\beq
\left\lbrace i \ga^{\al} \pa_{\al} - m
+ \frac{i}{2} \left(  h \eta^{\al \be}
- h^{\al \be} \right) \ga_{\be}\, \pa_{\al}
+ \frac{i}{4} \left( \pa_{\be} h
- \pa_{\al} h^{\al}_{\be} \right) \ga^{\be}
- \frac{m h}{2} \right\rbrace \psi \,=\, 0.
\label{eq3}
\eeq
Starting from this point, the parenthesis indicate (except where this
does not lead to a confusion) the limit of the action of derivative,
i.e., $\pa AB = (\pa A) B + A \pa B$. The next step is to separate
time $t$ from the space coordinates $x_i$ and denote
$\ga^0 = \be$, that gives
\beq
&&
\biggl\{ i\be \,\frac{\pa}{\pa t}
\,+\, i \ga^{k} \pa_{k} - m
- \frac{m h}{2} + \frac{i}{2} \big( h - h^{00} \big) \be \,\pa_t
+ \frac{i}{2} \big( h \ga^{k} \pa_{k} + h^{jk} \ga^{j} \pa_{k} \big)
\nn
\\
&&
\qquad
+\,\, \frac{i}{2} h^{0k} \ga^{k} \pa_{t}	
- \frac{i}{2} h^{k0} \be \pa_{k}
+ \frac{i}{4} \dot{h} \be
+ \frac{i}{4} \left( \pa_{k} h \right) \ga^{k}
- \frac{i}{4} \dot{h}^{00} \be
+ \frac{i}{4} \dot{h}^{0k} \ga^{k}
\nonumber
\\
&&
\qquad
-\,\,\frac{i}{4} \big( \pa_{k} h^{k0} \big)  \be
+ \frac{i}{4} \big( \pa_{k} h^{kj} \big) \ga^{j} \biggr\}\,
\psi \,\,=\,\, 0.
\label{eq4}
\eeq
Multiplying the last equation by $\be$, we change the notations
to $\be \ga^{k} = \al^{k}$, and use
\beq
h^{jk} = h_{jk},
\qquad
h^k_j = - h_{jk},
\qquad
h^{0k} = h^{k0} = - h_{k0},
\qquad
h_{0k} = - h^k_0,
\label{eq5}
\eeq
to arrive at
\beq
&&
\biggl\{
i \Big( 1 + \frac{1}{2} h - \frac{1}{2} h_{00}
- \frac{1}{2} h_{0k} \al^{k} \Big) \,\frac{\pa}{\pa t}
\,+\, i \Big( \al^{k} + \frac{1}{2} h \al^{k}
+ \frac{1}{2} h_{kj} \al^{j}
+ \frac{1}{2} h_{0k} \Big) \pa_{k}
\nn
\\
&&
\qquad
-\,\, \be m \Big( 1 + \frac{1}{2} h \Big)
\,+\, \frac{i}{4} \dot{h} - \frac{i}{4} \dot{h}_{00}
+ \frac{i}{4} \pa_{k} h_{k0}
+\, \frac{i}{4} \left( \pa_{k} h \right) \al^{k}
\nn
\\
&&
\qquad
-\,\, \frac{i}{4} \dot{h}_{0k}\, \al^{k}
+ \frac{i}{4} \left( \pa_{j} h_{jk} \right) \al^{k} \biggr\}
\psi \,=\, 0 .
\label{eq6}
\eeq
Starting from this point, the Latin indices are raised and lowered
using the metric of the Euclidean space.

The next operation is to ``liberate'' the time derivative. For this,
we remember that the components of $h_{\mu\nu}$ are taken
into account only up to the linear approximation and multiply
Eq.~ \eqref{eq6} by the factor
\beq
\Big(1 + \frac{1}{2} h - \frac{1}{2} h_{00}
- \frac{1}{2} h_{0k}\, \al^{k} \Big)^{-1}
\,=\,
1 - \frac{1}{2} h + \frac{1}{2} h_{00} + \frac{1}{2} h_{0k}\,\al^{k}.
\label{eq7}
\eeq
In this way, after a small algebra,	we get
\beq
&&
\biggl\{ i \,\frac{\pa}{\pa t} + i \Big( \al^{k}
+ \frac{1}{2} h_{00} \,\al^{k}
+ \frac{1}{2} h_{0j}\, \al^{j} \al^{k}
+ \frac{1}{2} h_{kj}\, \al^{j}
+ \frac{1}{2} h_{0k} \Big) \pa_{k}
\label{eq8}
\\
&&
\qquad
+\,\, \frac{i}{4} \Big(  \dot{h}
- \dot{h}_{00} + \pa_{k} h_{k0} \Big)
- \be m \Big( 1 + \frac{1}{2} h_{00}
+ \frac{1}{2} h_{0k}\, \al^{k} \Big)
\nonumber
\\
&&
\qquad
+\,\, \frac{i}{4} \big( \pa_{k} h \big) \al^{k}
- \frac{i}{4} \dot{h}_{0k}\, \al^{k}
+ \frac{i}{4} \big( \pa_{j} h_{jk} \big) \al^{k} \biggr\}\, \psi
\,\,=\,\, 0.
\nonumber
\eeq

At this point, we recover $c$ and $\hbar$, change to
the momentum representation in the space sector, and introduce the
electromagnetic potential $A^\mu = (\phi,\,\vec{A})$, which means
trading
\beq
i\pa_{t} \rightarrow  i\hbar \pa_{t} - e \phi,
\qquad
- i\pa_{k} \rightarrow - i\hbar \pa_{k} = p_k,
\qquad
p_{k} \rightarrow  c\Pi_{k} = c\left( p_{k} - \frac{e}{c} A_{k}\right).
\label{eq9}
\eeq
As a result, we get
\beq
&&
\biggl\{ i \hbar \pa_{t} - e\phi - c\left( \al^{k}
+ \frac{1}{2} h_{00} \al^{k}
+ \frac{1}{2} h_{0j} \al^{j} \al^{k}
+ \frac{1}{2} h_{kj} \al^{j}
+ \frac{1}{2} h_{0k} \right) \Pi_{k}
\nn
\\
&&
+\, \frac{i \hbar}{4} \left(  \dot{h}
- \dot{h}_{00} \right)
- \be mc^2\Big( 1 + \frac12 h_{00} + \frac{1}{2} h_{0k} \al^{k}\Big)
+ \frac{i \hbar c}{4} \left( \pa_{k} h + \pa_{j} h_{jk}\right) \al^{k}
\nonumber
\\
&&
+\, \dfrac{i\hbar}{4} \left( c \pa_{k} h_{k0}
+ \dot{h}_{0k} \al^k \right) \biggr\} \,\psi = 0.
\label{eq10}
\eeq
In the standard representation, one can easily obtain the relation
\beq
\al^j \al^k \,=\,
\de^{jk}\,+\,i \ep^{jkl}	\Si^l,
\qquad
\mbox{where}
\qquad
\vec{\Si} = \begin{pmatrix}
\vec{\si} & 0
\\
0 & \vec{\si}
\end{pmatrix}.
\label{eq11}
\eeq
As a result, Eq.~ (\ref{eq10}) can be written as
\beq
i\hbar\pa_t\psi \,=\, \hat{H} \psi,
\qquad
\mbox{where}
\qquad
\hat{H} = \hat{H}_\al + \hat{H}_\be
\label{eq16}
\eeq
and
\beq
&&
\hat{H}_\al \,=\,
c\Big( \Pi_{k}+\dfrac{1}{2}h_{00}\Pi_{k}
+ \dfrac{1}{2}h_{jk}\Pi_{j} \Big) \al_k
- \dfrac{i\hbar c}{4}(\pa_k h +\pa_jh_{jk})\al^k
+ \dfrac{i\hbar}{4} \dot{h}_{0k} \al_k\,,
\label{eq18}
\\
&&
\hat{H}_\be
\,=\,
e\phi + \be m c^2 \Big(1+\dfrac{1}{2}h_{00}
+ \dfrac{1}{2}h_{0k}\al^k \Big)
\nn
\\
&&
\qquad
\qquad
+ \,\dfrac{i\hbar}{4}(\dot{h}_{00} - \dot{h})
+ \dfrac{ic}{2} h_{0j} \ep^{jkl} \Si^l \Pi_{k}
+ c \, h_{0k} \Pi_k
- \dfrac{i\hbar c}{4} \pa_k h_{k0}\,.
\label{eq17}
\eeq

Let us impose the gauge fixing condition, which helps to simplify
the expressions. The simplest option is $h_{0k}=0$, which gives
\beq
\hat{H}' \,=\,\hat{H}'_\al + \hat{H}'_\be,
\eeq
where
\beq
&&
\hat{H}'_\be
\,=\, e\phi +\be m c^2 \Big(1+\dfrac{1}{2}h_{00}\Big)
+ \dfrac{i\hbar}{4}(\dot{h}_{00}-\dot{h})
\label{eq21}
\\
&&	
\hat{H}'_\al
\,=\, c\Big( \Pi_{k}+\dfrac{1}{2}h_{00}\Pi_{k}
+\dfrac{1}{2}h_{jk}\Pi_{j} \Big) \al_k
- \dfrac{i\hbar c}{4}(\pa_k h+\pa_jh_{jk})\al^k
\,=\,c\al^kV_k.
\label{eq22}
\eeq
In the last formula, we introduced the following notations:
\beq
&&
V_k \,=\, \Big( \de_{jk} + \dfrac{1}{2}\om_{jk} \Big) \Pi_{j} + v'_k,
\label{eq23}
\\	
&&
v'_k \,=\,
- \dfrac{i\hbar}{4}(\pa_kh+\pa_jh_{jk}),
\qquad
\om_{jk}=h_{00}\de_{jk}+h_{jk}.
\label{eq24}
\eeq

As a first step to take the non-relativistic limit, we change
the variable
\beq
\psi\,=\,\psi' \exp \Big(-\dfrac{imc^2}{\hbar}t\Big).
\label{eq25}
\eeq
The new spinor $\psi'$ obeys the equation
\beq
i\hbar\pa_t\psi'
\,=\,
\Big\{
e\phi + (\be-I)mc^2+\dfrac{mc^2}{2}\be h_{00}
\,+\, \dfrac{i\hbar}{4}(\dot{h}_{00} - \dot{h})+H'_\al
\Big\}\psi'.
\label{eq27}
\eeq
In the standard representation
\beq
\be = \begin{pmatrix} I & 0 \\ 0 & -I \end{pmatrix},
\qquad
\vec{\al}\,=\,\begin{pmatrix} 0 & \vec{\si} \\ \vec{\si} & 0 \end{pmatrix},
\qquad
\mbox{and assuming}
\qquad
\psi'=\begin{pmatrix} \ph \\ \ch	\end{pmatrix},
\label{eq28}
\eeq
we meet two equations for the two-components spinors
$\ph$ and $\chi$,
\beq
&&
i\hbar\,\frac{\pa \ph}{\pa t}
\,-\,e\phi\ph - \dfrac{1}{2}mc^2h_{00}\ph
- \dfrac{i\hbar}{4}\big(\dot{h}_{00}-\dot{h}\big)\ph
\,=\,
c\vec{\si}\cdot\vec{V}\ch	
\label{eq30a}
\\
&&
i\hbar\,\frac{\pa \ch}{\pa t}
\,-\,
e\phi\ch+2mc^2\Big(1+\dfrac{1}{4}h_{00}\Big)\ch
- \dfrac{i\hbar}{4}\big(\dot{h}_{00}-\dot{h}\big)\ch
\,=\,
c\vec{\si} \cdot \vec{V}\ph.
\label{eq30b}
\eeq
In the left hand side of (\ref{eq30b}), the term
$2mc^2\big(1+h_{00}/4\big)$ is dominating in the
nonrelativistic regime. Thus, we can neglect other terms and
arrive at the approximate solution for the ``small'' component
of the Dirac spinor,
\beq
\ch \,=\, \Big(1-\dfrac{1}{4}h_{00}\Big)
\,\dfrac{\,\,\vec{\si}\cdot\vec{V}}{2mc}\,\ph.
\label{eq31}
\eeq
Substituting this result in Eq.~(\ref{eq30a}), we get
\beq
&&
i \hbar\,\frac{\pa \ph}{\pa t}
\,-\, e\phi\ph \,-\, \dfrac{mc^2}{2}h_{00}\ph
\,-\, \dfrac{i\hbar}{4} \big(\dot{h}_{00}-\dot{h}\big)\ph
\nn
\\
&&
\quad
\,=\,
\dfrac{1}{2m}
\Big(1-\dfrac{1}{4}h_{00}\Big)(\vec{\si}\cdot\vec{V})^2 \ph
+ \dfrac{i\hbar}{8m} (\pa_k h_{00}) \Pi_k\,\ph
- \dfrac{\hbar}{8m} \si^j \ep^{jkl} (\pa_k h_{00}) \Pi_l\, \ph,
\qquad
\label{eq32}
\eeq
where the first term in the right hand side requires an additional 
linearization in
$h_{\mu\nu}$ components, according to Eqs.~(\ref{eq23})
and (\ref{eq24}). After this, Eq.~(\ref{eq32}) becomes
a non-relativistic approximation to the ``large'' component
of the Dirac spinor, in the presence of electromagnetic and weak
gravitational fields. The last enters this expression in a rather
complicated way, as one can see from the expression for $\vec{V}$
in \eqref{eq23}.

The subsequent transformations are pretty much standard, including
\beq
(\vec{\si}\cdot\vec{V})^2
\,=\,
\dfrac{1}{2}(\si^j\si^k+\si^k\si^j)V^jV^k
+ \dfrac{1}{2}(\si^j\si^k-\si^k\si^j)V^jV^k
\,=\,  V_kV_k + i\si^l\epsilon^{ljk}V_jV_k,
\label{eq33}
\eeq
where $V_kV_k = \vec{V}^2$. The \textit{r.h.s.} of Eq.~\eqref{eq32}
becomes
\beq
&&
\bigg\{\dfrac{1}{2m}\,
\Big(1-\dfrac{1}{4}h_{00}\Big)\,V_kV_k
+ \dfrac{i}{2m}\,\Big(1-\dfrac{1}{4}h_{00}\Big)\si_k\epsilon^{klj}\,V_lV_j
\nn
\\
&&
\qquad \qquad
\,+\,\dfrac{i\hbar}{8m} (\pa_k h_{00}) \Pi_k
- \dfrac{\hbar}{8m}\, \si^j \ep^{jkl} (\pa_k h_{00})\, \Pi_l
\bigg\}\ph.
\label{eq34}
\eeq
Regardless these formulas may be boring, let us elaborate this part
in detail. For the first term, we have
\beq
\dfrac{1}{2m}\,\Big(1-\dfrac{1}{4}h_{00}\Big)\,V_kV_k
&  =  &
\dfrac{1}{2m}\,\Big(1-\dfrac{1}{4}h_{00}\Big)
{\vec{\Pi}}^2
\,-\, \dfrac{i\hbar}{4m}\,
\left[ (\pa_k h_{00} ) + (\pa_i h_{ik}) \right] \Pi_k
\nn
\\
&&
+\,\,\dfrac{1}{2m}\om_{kl}\Pi_k \Pi_l \,+\,\dfrac{1}{m}v'_k\Pi_k
\label{eq35}	
\eeq
and for the second term,
\beq
\frac{i}{2m} \Big( 1 - \frac{1}{4}	h_{00} \Big)
\vec{\si} \cdot \big[ \vec{V} \times \vec{V} \big]
&=&
-\,\dfrac{e \hbar}{2mc}
\Big( 1 - \frac{1}{4}	h_{00} \Big) \vec{\si} \cdot  \vec{B}
\,+ \,\dfrac{\hbar}{4m} \,\si_{m} \ep^{mkl}
	\left( \pa_{k} \om_{lr} \right)  \Pi_{r}
\nonumber
\\
&&
+ \,\dfrac{i}{4m} \si_{m} \epsilon^{mkl}
\om_{lr} \left( \Pi_{k} \Pi_{r} - \Pi_{r} \Pi_{k} \right),
\nn
\eeq
where $ \vec{B} $ is a magnetic field. In the last equation, we have
\beq
\Pi_{k}\Pi_{r}-\Pi_{r}\Pi_{k}
\,=\, \dfrac{ie\hbar}{c} F_{kr}
\,=\, \dfrac{ie\hbar}{c} \epsilon_{jrk} B_j .
\label{eq37}
\eeq
Finally, Eq.~\eqref{eq32} becomes
\beq
i \hbar\,
\frac{\pa  \ph}{\pa t}
&=&
\biggl\{
e \phi + \dfrac{1}{2} mc^{2} h_{00}
+ \dfrac{i \hbar}{4} \big( \dot{h}_{00} - \dot{h} \big)
\,-\, \dfrac{i\hbar}{4m} \left[\dfrac{1}{2}\, (\pa_k h_{00})
+ (\pa_i h_{ik}) \right] \Pi_k
\nn
\\
&&
+ \,\, \dfrac{1}{2m} \Big( 1 - \dfrac{1}{4} h_{00} \Big)
\vec{\Pi}^2 - \dfrac{e \hbar}{2mc} \Big( 1 - \frac{1}{4}	h_{00} \Big)
\vec{\si} \cdot  \vec{B} + \dfrac{1}{2m} \om_{kl} \Pi_{k} \Pi_{l}
\nn
\\
&&
+ \,\, \dfrac{1}{m} v'_{k} \Pi_{k} +
\dfrac{\hbar}{4m} \si_{m} \epsilon^{mkl}
\left( \pa_{k} \tilde{\om}_{lr} \right) \Pi_{r}
- \dfrac{e\hbar}{4mc} \si_{m} \epsilon^{mkl}
\epsilon_{jrk} \om_{lr} B_j \biggr\}\, \ph,
\label{eq38}
\eeq
where we defined a new parameter
$\tilde{\om}_{rl}=(\frac{1}{2}h_{00}\de_{lr}+h_{lr})$.
This is the extended version of the Pauli equation, taking into
account the gravitational field $h_{\mu\nu}$ restricted by the
gauge fixing  condition $h_{0k}=0$. Setting $h_{\mu\nu}=0$,
we arrive at the usual version of this equation \cite{LL-4}.

We can make the following observation. In some
cases, e.g., for a field corresponding to a nonrelativistic massive
particle with spin moving in an external gravitational field, the
components of $h_{\mu\nu}$ may be proportional to the mass
$m$ of the test particle. In such a situation, since the expansion is
in the inverse powers of $m$, one can expect ambiguities and
scheme-dependent results, as
it was discussed in \cite{Obukhov-2000} and other references.
We shall discuss examples of such an ambiguity below.

\section{Equations of motion for a non-relativistic particle}
\label{s4}

The result of the previous section is Eq.~(\ref{eq38}), which
can be rewritten in the form
\beq
i \hbar\,\frac{\pa  \ph}{\pa t} &=& \hat{H} \ph,
\label{eq39}
\eeq
where
\beq
\hat{H} \,=\,
A - \vec{\si} \cdot \vec{W} + C_{kl} \Pi_{k} \Pi_{l} + D_{k} \Pi_{k}.
\label{eq40}
\eeq
In the \textit{r.h.s.} of this formula and below we omit the hats for
the sake of brevity. The elements of (\ref{eq40}) are defined as follows
\beq
&&
A \,=\, e\phi + \dfrac{1}{2}mc^{2} h_{00}
+ \dfrac{1}{4}i \hbar \big( \dot{h}_{00} - \dot{h} \big),
\nn 
\\
&&
W^i = \dfrac{e\hbar}{2mc}\,
\Big( 1 - \frac{1}{4}h_{00} \Big) B^i
\,+\, \dfrac{e\hbar}{4mc}\, \epsilon^{ikl} \epsilon_{jrk}\,
\om_{lr}\, B_j \,, 
\nn 
\\
&&
C_{kl} = \dfrac{1}{2m} \de_{kl} \Big( 1 - \frac{1}{4}h_{00} \Big)
+ \dfrac{1}{2m} \om_{kl} ,
\nn 
\\
&&	
D_{k} = \dfrac{1}{m} v'_{k} + \dfrac{\hbar}{4m} \si_{m} \ep^{mrl}
\left( \pa_{r} \tilde{\om}_{lk} \right)
\,-\,
\dfrac{i\hbar}{4m} \Big[ \dfrac{1}{2}\,(\pa_k h_{00} )
+ (\pa_i h_{ik}) \Big] \,.
\label{eq44}
\eeq

The four elements (\ref{eq44})
depend on time and also on the canonical operators, i.e., coordinates
$x_i$, momenta $p_j$, and spin $\si_k$ components. These operators
satisfy canonical commutation relations, which are all zero except
\beq
\left[ x_i,p_j \right] =  i\hbar \de_{ij}
\qquad \mbox{and} \qquad
\left[ \si_{i},\si_{j}\right] = 2i \epsilon_{ijk} \si_k \,.
\label{can2}
\eeq
The classical limit is obtained from the Heisenberg equation.
For the operator $\mathcal{O}(t)$,
\beq
\frac{d \mathcal{O}}{dt}
&=& \frac{1}{i\hbar}\,\big[\mathcal{O},H\big]
\label{Heiseq}
\eeq
should be taken in the limit $\hbar \to 0$. In what follows, we
shall use this scheme to derive the equations of motion for classical
canonical variables $x_i$, $p_j$  and  $\si_k $.

\subsection{Equation of motion for the coordinates}

In the case of $x_i$, we need to evaluate the commutator
\beq
\left[ x_i, H \right]
&=& \left[ x_i, C_{kl}\Pi_{k}\Pi_{l} \right]
\,+\,  \left[ x_i, D_{k}\Pi_{k} \right].
\label{eq46}
\eeq
At this point, things start to simplify. Since
$ D_{k} =\mathcal{O}(\hbar)$, we have
$\left[ x_i, D_{k}\Pi_{k} \right]=\mathcal{O}(\hbar^2)$.

Since we are interested in the limit $\,\hbar \rightarrow 0$, and
taking into account Eq.~(\ref{Heiseq}), this term is irrelevant. In
the remaining terms, also omitting all $\mathcal{O}(\hbar^2)$
contributions, we get
\beq
\left[ x_i, C_{kl}\Pi_{k}\Pi_{l} \right]
&=&
\dfrac{1}{2m} \left[ \Big(1 - \dfrac{1}{4} h_{00}\Big) \de_{kl}
+ \om_{kl} \right] \big[ x_i, \Pi_{k}\Pi_{l} \big]
\nn
\\
&&
=\,\,
\dfrac{i\hbar}{m}\left[ \Big(1 + \dfrac{3}{4} h_{00}\Big)\Pi_i
+  h_{il} \Pi_{l} \right].
\label{eq49}
\eeq
In the limit $\hbar \rightarrow 0$, the equation of motion for
$x_i$ becomes
\beq
\dfrac{dx_{i}}{dt}\, = \,
\dfrac{1}{m}\left[ \Big(1 + \dfrac{3}{4} h_{00}\Big)\Pi_{i}
+  h_{il} \Pi_{l} \right].
\label{eq51}
\eeq

\subsection{Equation of motion for the momentum}

To obtain the equations of motion for $p_i$, we note that
$\vec{W} = \mathcal{O}(\hbar)$ and $D_k = \mathcal{O}(\hbar)$.
For this reason, $\left[p_i,\vec{W} \right]= \mathcal{O}(\hbar^2)$
and $\left[p_i,D_k\right]= \mathcal{O}(\hbar^2)$, Thus, the
corresponding terms can be neglected in the limit
$\hbar \rightarrow 0$ in Eq.~(\ref{Heiseq}). The relevant
commutator for $p_i$ is
\beq
\left[p_i,H \right] = \left[p_i, A \right]
+ \left[p_i, C_{kl} \Pi_{k} \Pi_{l} \right].
\label{eq52}
\eeq
For the remaining terms, we get
\beq
&&
\left[p_i, A \right]
\,=\,
- ie\hbar \pa_i \phi - \dfrac{i \hbar}{2} mc^2
\left( \pa_i h_{00} \right) + \mathcal{O}(\hbar^2),
\label{eq53}
\\
&&
\left[ p_i, C_{kl} \Pi_{k} \Pi_{l} \right]
\,=\,
\left[ p_i, C_{kl} \right]
\Pi_{k} \Pi_{l} + C_{kl} \left[ p_i,  \Pi_{k}  \right] \Pi_{l}
+ C_{kl} \Pi_{k} \left[ p_i,  \Pi_{l} \right],
\label{eq54}
\eeq
where
\beq
\left[ p_i, C_{kl} \right] \,=\, -\dfrac{i\hbar}{2m}
\Big[ \dfrac{3}{4} \de_{kl} (\pa_{i} h_{00})
+ (\pa_{i} h_{kl})\Big]
\qquad
\mbox{and}
\qquad
\left[ p_i,  \Pi_{k}  \right]  \,=\, \dfrac{ie\hbar}{c} (\pa_{i} A_k).
\label{eq55}
\eeq
Substituting \eqref{eq55} in \eqref{eq54}, we arrive at
\beq
\left[ p_i, C_{kl} \Pi_{k} \Pi_{l} \right]
&=& -\,\, \dfrac{i\hbar}{2m}
\left[ \dfrac{3}{4} \de_{kl} (\pa_{i} h_{00})
+ (\pa_{i} h_{kl})\right] \Pi_{k} \Pi_{l}
\label{eq56}
\\
&&
+ \,\,\dfrac{ie\hbar}{c} C_{kl} (\pa_{i} A_k) \Pi_{l}
+ \dfrac{ie\hbar}{c} C_{kl} \Pi_{k} (\pa_{i} A_k).
\nonumber
\eeq
Finally, using \eqref{eq53} and \eqref{eq56} in Eq.~\eqref{eq52},
we obtain
\beq
\left[p_i,H \right]
&=& - \, \, i\hbar \left[ e \pa_i \phi
+ \dfrac{mc^2}{2} \left( \pa_i h_{00} \right) \right]
- \dfrac{i\hbar}{2m} \left[ \dfrac{3}{4} \de_{kl} (\pa_{i} h_{00})
+ (\pa_{i} h_{kl})\right] \Pi_{k} \Pi_{l}
\label{eq57}
\\
&&
+\,\, \dfrac{ie\hbar}{c}
\left[ C_{kl} (\pa_{i} A_k) \Pi_{l}
+ C_{kl} \Pi_{k} (\pa_{i} A_k)\right].
\nonumber
\eeq

Next, taking the limit $\hbar \rightarrow 0$, we arrive at the
classical equation of motion for $p_i$
\beq
\dfrac{dp_i}{dt} \,=\, eE_i
- \dfrac{mc^2}{2} \left( \pa_i h_{00} \right)
+ \dfrac{2e}{c} C_{kl} (\pa_{i} A_k) \Pi_{l}
- \dfrac{1}{2m} \left[ \dfrac{3}{4} \de_{kl} (\pa_{i} h_{00})
+ (\pa_{i} h_{kl})\right] \Pi_{k} \Pi_{l},
\label{eq58}
\eeq
where $ E_i = -\pa_i \phi $.
Now we remember that $ \Pi_i = p_i - \frac{e}{c}A_i $, thus
\beq
\dfrac{dp_i}{dt} \,=\, \dfrac{d \Pi_i}{dt}
+ \dfrac{e}{c}\dfrac{\pa A_i}{\pa x_k} \dfrac{dx_k}{dt}
+ \dfrac{e}{c} \dfrac{\pa A_i}{\pa t}.
\label{eq59}
\eeq
Substituting Eq.~\eqref{eq51} in \eqref{eq59}, we get
\beq
\dfrac{dp_i}{dt} \,=\,  \dfrac{d\Pi_i}{dt}
+ \dfrac{e}{mc}\dfrac{\pa A_i}{\pa x_k}
\left[ \left( 1 + \dfrac{3}{4} h_{00} \right)\Pi_{k}
+  h_{kl} \Pi_{l} \right] + \dfrac{e}{c}\dfrac{\pa A_i}{\pa t}.
\label{eq60}
\eeq
Finally, substituting \eqref{eq60} into \eqref{eq58}, we arrive at
the result
\beq
\dfrac{d\Pi_i}{dt} &=& eE_i - \dfrac{mc^2}{2} (\pa_{i} h_{00})
- \dfrac{1}{2m} \left[ \dfrac{3}{4} \de_{kl} (\pa_{i} h_{00})
+ (\pa_{i} h_{kl}) \right] \Pi_{k}\Pi_{l}
\nn
\\
& & + \dfrac{e}{mc}
\left( \de_{kl} + \frac{3}{4} \de_{kl} h_{00} + h_{kl} \right)
\Pi_{k} \epsilon_{ilj}B_j.
\label{eq61}
\eeq

\subsection{Equation of motion for the spin}

For the spin variable $ \si_{i} $, the initial commutator has the form
\beq
\left[ \si_{i}, H \right]
\, = \,
\left[ \si_{i}, \,- \vec{\si} \cdot \vec{W}
+ \dfrac{\hbar}{4m} \si_{m} \epsilon^{mrl}
\left( \pa_{r} \tilde{\om}_{ln} \right) \Pi_{n} \right].
\label{eq62}
\eeq
Using the commutation relation (\ref{can2}), for the sigma
matrices, we have, after some algebra,
\beq
\left[ \si_{i}, H \right]
&=&
2i \epsilon_{ijk} \si_k
\biggl\{-\dfrac{e\hbar}{2mc}
\Big( 1 - \dfrac{1}{4} h_{00} \Big) B_{j}
- \dfrac{e\hbar}{4mc} \epsilon^{jml} \om_{lr}
\ep_{nrm} B_n
\nn
\\
&&
+ \,\dfrac{\hbar}{4m} \,
\epsilon^{jml} \left( \pa_{m} \tilde{\om}_{ln} \right)
\Pi_{n}  \biggr\}.
\nonumber
\\
&=&
-\dfrac{ie\hbar}{mc} \left( 1 - \dfrac{1}{4} h_{00} \right)
\epsilon_{ijk} B_{j} \si_k + \dfrac{ie\hbar}{2mc}
\epsilon_{ijk} \left( \om_{rr} B_j - \om_{lj} B_l \right) \si_k
\label{eq66}
\\
&&
- \dfrac{i\hbar}{2m} \,
\big[
\left( \pa_{i} \tilde{\om}_{kn} \right)
- \left( \pa_{k} \tilde{\om}_{in} \right) \big] \Pi_{n} \si_k.
\nonumber
\eeq
Then, in the limit $\hbar \rightarrow 0$, we get the equation
of motion for $\si_i$,
\beq
\dfrac{d \si_{i}}{dt}
&=&
-\dfrac{e}{mc} \,
\Big( 1 - \dfrac{1}{4} h_{00} \Big) \,\epsilon_{ijk} B_{j} \si_k
+ \dfrac{e}{2mc}\,
\epsilon_{ijk} \big( \om_{rr} B_j - \om_{lj} B_l \big) \si_k
\nn
\\
&&
- \,\,
\dfrac{1}{2m} \big[
\left( \pa_{i} \tilde{\om}_{kn} \right)
- \left( \pa_{k} \tilde{\om}_{in} \right) \big]
\Pi_{n} \si_k.
\label{eq67}
\eeq

The last observation is that all three equations (\ref{eq51}),
(\ref{eq61}), and (\ref{eq67}) may be incomplete in the case
when $h_{\mu\nu} \sim m$, as we explained in the previous
section.

\section{The special case of the plane gravitational wave background}
\label{s5}

The equations (\ref{eq51}), (\ref{eq61}) and (\ref{eq67}) obtained
in the previous section are not the most general ones because of the
gauge fixing condition $h_{0k}=~0$. It is worth
discussing the physical sense of this restriction. In gravity, the
gauge transformations are related to the coordinate transformations.
Fixing the gauge, we choose a reference frame, as it happens in
cosmology with the synchronous gauge fixing. However, in the
present case, we have no right to perform the transformations of
coordinates in the expansion (\ref{hmn}) because the non-inertial
coordinates may result in the inertial force being equivalent to
the gravitational field.
In particular, this means that our results do not admit consideration
of the gravitational field of rotating body. On the other hand, these
results are sufficient to consider the two important examples, namely
the plane gravitational wave  on a flat background and the
Schwarzschild metric. Let us start from the first of these two cases.

Consider the metric of the weak plane gravitational wave propagating
along the axis $OX$. The  nonzero components of the gravitational
perturbations are
\beq
h_{yy} = - h_{zz} = -2\up,
\qquad
h_{yz} = h_{zy} = 2u,
\label{eq68}
\eeq
where $\up$ and $u$ are the functions of the light cone coordinate
$ct - x$. Each of these functions describes one possible polarization
of the gravitational wave.	To compare our results with some of those
available in literature, consider only one polarization state, i.e.,
set $u=0$. The relevant quantities are
\beq
\tilde{\om}_{kl} = \om_{kl} = 2\up T_{kl}
\qquad
\mbox{and}
\qquad
v'_{k} = - \,\dfrac{i \hbar}{2} T_{jk} (\pa_{j} \up),
\label{eq69}
\eeq
where we introduced the notation
\beq
T_{jk} \,=\,
\begin{pmatrix} 0 & 0 & 0 \\ 0 & -1 & 0 \\ 0 & 0 & 1 \end{pmatrix}\,.
\label{eq70}
\eeq
Now we are in a position to apply the general formulas to
the plane wave background.

\subsection{Pauli equation on the  background of a
gravitational	wave}

Using (\ref{eq69}) in Eq.~\eqref{eq38}, after some algebra,
we arrive at the equation
\beq
i \hbar \,\frac{\pa  \ph}{\pa t}
&=&
\biggl\{ e \phi + \dfrac{1}{2m}\Pi_k \Pi_k
- \dfrac{e \hbar}{2mc} \vec{\si} \cdot  \vec{B}
+ \dfrac{v}{m} T_{kl} \Pi_{k} \Pi_{l}
\nn  
\\
&&
+ \dfrac{\hbar}{2m} \si_{m} \epsilon^{mkl} T_{lr}
\Pi_{r} \left( \pa_{k} v \right) - \dfrac{e \hbar v}{2 mc}
\left( \si_{m} T_{ll} B_{m} - \si_{r} T_{jr} B_{j} \right)
\biggr\} \ph
\nn
\\
&=& \biggr\{ e \phi + \dfrac{1}{2m} \left( \de_{kl}
+ 2v T_{kl} \right) \Pi_{k} \Pi_{l}
- \dfrac{e \hbar}{2mc} \left( \de_{kl}
+ v T_{kl} \right) \si_{k} B_{l}
\nn
\\
&&
+ \dfrac{\hbar}{2m} \si_{m} \epsilon^{mkl} T_{lr} \Pi_{r}
( \pa_{k} v )\biggr\} \ph .
\label{eq72}
\eeq
This equation can be compared with the one obtained by Quach
in \cite{Quach-2015} (which made a correction to the previous
publication \cite{Goncalves-2007}), using both perturbative and
exact Foldy-Wouthuysen transformations. Using the units with
$\hbar=c=e=1$, our Hamiltonian becomes
\beq
H
&=&
m + \dfrac{1}{2m} \left( \de_{kl}
+ 2\up T_{kl}\right)\Pi_{k}\Pi_{l}
- \dfrac{1}{2m} \left( \de_{kl}
+ \up T_{kl} \right) \si_{k} B_{l}
\nn
\\
&&
+ \,\,\dfrac{1}{2m} \si_{m} \epsilon^{mkl}
T_{lr} \Pi_{r} \left( \pa_{k} \up \right).
\label{eq73}
\eeq
Furthermore, we elaborate some expressions as follows:
\beq
&&
\dfrac{\be}{4m}\,\de^{ij}\ep_{jkl}\,\si^l(\pa^kA_i-\pa_iA^k)
\,=\,-\,\dfrac{\be}{2m}\Si^lB_l,
\,\quad\quad
\label{eq76}
\eeq
where we used $\ep_{jkl}\pa^k A_i=\de_{il}B_j$, and also
\beq
\dfrac{\be}{4m}
T^{ij}\,\ep_{jkl}\,\Si^l\,(\pa^kA_i-\pa_iA^k)
&=& -\,\,\dfrac{\be}{2m}\,
\Si^i\,T^{ij}\,B_j.
\label{eq77}
\eeq
Using these relations, taking into account the definition of
$\vec{\Si}$ in (\ref{eq11}) and applying the Hamiltonian of
\cite{Quach-2015} to the wave function $\psi '$ in the form
(\ref{eq28}), the result is identical to (\ref{eq73}), except the
factor 2 which multiplies $\up$ of the last term in the first line
of Eq. (\ref{eq73}). We did not find this factor in our calculation.
This difference is likely a misprint, which does not affect the
equations of motion for the spinning particle found in \cite{Quach-2015}.  With this small exception, our
general formula and the Pauli equation on the background of a
plane gravitational wave~\cite{Quach-2015}, got confirmed.

\subsection{Equations of motion for a nonrelativistic particle}

Let us start with the equations for coordinates. Starting from
Eq.~(\ref{eq51}), we use the metric of the gravitational wave,
as explained above. The result is
\beq
\dfrac{dx_{i}}{dt} \,=\,
\dfrac{1}{m}
\left( \de_{il} + 2v T_{il} \right) \Pi_{l},
\label{eq81}
\eeq
that is equivalent to the corresponding equation in
\cite{Goncalves-2007}.

Consider the equations of motion for the momenta components
$ p_i $. Using the gravitational wave metric from Eq.~(\ref{eq58}),
we arrive at
\beq
\dfrac{dp_i}{dt}
\,\,=\,\,
eE_i \,-\, \dfrac{1}{m}\, T_{kl}
(\pa_i \up)\,\Pi_{k} \Pi_{l}
\,+\, \dfrac{e}{mc}
\big(\de_{kl} + 2\up T_{kl} \big) \,\Pi_{k} \big(\partial_i A_l\big).
\label{eq85}
\eeq
The unique difference with the analogous equation in
\cite{Goncalves-2007} is the term with electric field, which
was not discussed in this reference.

Finally, for the spin components $ \si_{i}$, the equations of
motion can be derived by using the gravitational wave metric
in Eq.~(\ref{eq67}), that gives
\beq
\dfrac{d\sigma_{i}}{dt}
\,=\,
- \, \dfrac{e}{mc} \,\ep_{ijk} B_{j} \sigma_{k}
\,-\, \dfrac{e\,\up}{mc} \, \ep_{ijk} \,T_{lj}\, B_l \sigma_k
\,-\, \dfrac{1}{m} \big[ \left( T_{kn} \partial_{i} \upsilon \right)
\,-\, \left( T_{in} \partial_{k} \upsilon \right) \big] \Pi_{n} \sigma_k.
\label{sigmaGW}
\eeq
This equation is different from the one in \cite{Goncalves-2007}.
However, by using the relation
\beq
\label{eq1043}
F_{ln} \,=\, \partial_{l} A_n - \partial_n A_l
\,=\, - \,\,\ep_{nlr}\, B_r,
\eeq
it can be shown equivalent to the formula obtained by Quach in
\cite{Quach-2015}, except the sign of the first, non-gravitational
term, which is correct in (\ref{sigmaGW}).

\section{Newtonian limit in the Schwarzschild's metric}
\label{s6}

One of the simplest ways to arrive at the constant Newtonian
gravitational field is to consider a weak regime of the spherically
symmetric case. As it was discussed above, the nonrelativistic limit
in this case is expected to produce complications because the
potential energy is proportional to the mass of the test particle.
Let us explore this situation in detail.

The Schwarzschild's metric tensor is given by
\beq
&&
g_{\mu \nu} \,=\, \diag \big(A,\,\, - A^{-1},\,\, -r^2
,\,\, -r^2 \sin^2\th \big),
\nn
\\
&&
\qquad
\,=\, \diag \big(A,\,\, - \tilde{A},\,\, -r^2
,\,\, -r^2 \sin^2\th \Big),
\label{eq92}
\eeq
where
\beq
A \,=\,A(r)\,=\,1 - \dfrac{2GM}{rc^2}\,,
\qquad
\tilde{A}\,=\,\tilde{A}(r)\,=\,1 + \dfrac{2GM}{rc^2}
\,=\,A^{-1}(r)
\label{Apm}
\eeq
and we used the weak-field limit, $2GM / rc^2 \ll 1$ for
a weak gravitational field of a point-like mass $M$.
Changing to the isotropic coordinates,
\beq
r \,=\, r'\Big( 1 \,+\, \dfrac{GM}{2 r'c^2} \Big)^2\,,
\label{eq93}
\eeq
Eq. ~\eqref{eq92} becomes, after omitting the prime
\beq
g_{\mu \nu} \,=\, \diag \big(A,\,\, - \tilde{A},\,\,
-\, r^2\tilde{A},\,\, -\,r^2 \tilde{A} \sin^2\th  \big).
\label{eq94}
\eeq
In the  new (quasi-cartesian) coordinates
\beq
x = r \cos \phi \sin \th,
\quad
y = r \sin \phi \sin \th,
\quad
z = r \cos \th,
\eeq
the metric is cast as
\beq
g_{\mu \nu} \,=\, \diag \big(A,\,\, - \,\tilde{A},\,\,
-\, \tilde{A},\,\, -\, \tilde{A} \sin^2\th  \big),
\label{eq95}
\eeq
that is (\ref{hmn}), \
$g_{\mu \nu} = \eta_{\mu \nu} + h_{\mu \nu}$, with
\beq
h_{\mu \nu} \,= \, \dfrac{2\Phi}{c^2} \,
\diag \big(1,\,1,\,1,\,1\big),
\qquad
\mbox{where}
\qquad
\Phi \,=\, -\,\dfrac{GM}{r}
\label{eq98}
\eeq
is the Newtonian potential. The relevant summands of the
vector $V_k$ in \eqref{eq24} are
\beq
\om_{kl} \,=\, \dfrac{4 \Phi}{c^2} \de_{kl},
\qquad
\tilde{\om}_{kl}=\dfrac{3\Phi}{c^2}\de_{kl},
\qquad
v'_{k} \,=\, \dfrac{i \hbar}{2c^2} \,(\pa_{k} \Phi)\,.
\label{eq99}
\eeq
Now we are in a position to consider the Pauli equation and
the equations of motion for the particle. The results should be
compared with the ones of Ref.~\cite{Silenko-2005}.

\subsection{Pauli equation in the Newtonian limit}
\label{61}

Introducing the metric given by Eq.~\eqref{eq98} and the terms
\eqref{eq99} in Eq.~\eqref{eq38}, we get, for the Newtonian limit
of the Pauli-like equation,
\beq
i \hbar \,
\frac{\pa \ph}{\pa t}
&=&
\biggl\{ e\phi + m \Phi
+ \dfrac{1}{2m} \Big( 1 + \dfrac{7 \Phi}{2c^2} \Big) \, \vec{\Pi}^2
- \dfrac{e\hbar}{2mc} \Big(1 + \dfrac{9\Phi}{2c^2} \Big)\,
\vec{\si} \cdot \vec{B}
\label{eq100}
\\
&&
-\,\,\dfrac{i\hbar}{4mc^2} \left( \pa_{k} \Phi \right) \Pi_{k}
+ \dfrac{3\hbar}{4mc^2} \,\si_{m} \epsilon^{mkl} \left( \pa_{k} \Phi
\right) \Pi_{l}\biggr\} \ph.
\nonumber
\eeq
To compare with the corresponding equation of \cite{Silenko-2005},
we have to take zero electromagnetic field, use the natural system of
units $\hbar=c=e=1$ and add the rest energy $mc^2 = m$. With these
changes, the Hamiltonian operator in (\ref{eq100}) becomes
\beq
H\,=\,m + m\Phi
+ \dfrac{1}{2m} \Big(1 +\dfrac{7}{2}\Phi\Big)p^2
- \dfrac{i}{4m}\,(\pa_k\Phi)\,p_k
+ \dfrac{3}{4m}\,\si^j\ep^{jkl}(\pa_k\Phi)\,p_l\,.
\label{eq101}
\eeq
On the other hand, the Hamiltonian obtained in \cite{Silenko-2005}
has the form
\beq
H_{ST}\,=\,
\be\ep-\dfrac{\be}{2}\left\{\dfrac{\ep^2+p^2}{\ep},\dfrac{GM}{r}\right\}
\,\,-\,\,
\dfrac{\be\,(2\ep+m)}{4\ep(\ep+m)}\,
\big[ 2\vec{\Si} \cdot(\vec{g} \times \vec{p}) \,+\, \na \vec{g} \big],
\label{eq102}
\eeq
where
\beq
\{A,B\}=AB+BA,
\qquad
\ep=\sqrt{m^2+p^2},
\qquad
\vec{g} \,=\, - \, \na\Phi,
\label{eq103a}
\eeq
and $\vec{\Si}$ has been defined in (\ref{eq11}).
Looking for the correspondence with (\ref{eq101}), we expand
\beq
\ep
\,=\,
m\sqrt{1+\dfrac{p^2}{m^2}}
\,\simeq \,
m\Big(1+\dfrac{p^2}{2m^2}\Big).
\label{eq104a}
\eeq
In the same approximation, we get
\beq
\dfrac{\ep^2+p^2}{\ep}\,\simeq \,
m \Big(1+\dfrac{3p^2}{2m^2}\Big),
\qquad
\dfrac{2\ep+m}{\ep(\ep+m)}\simeq \dfrac{3}{2m},
\eeq
and (\ref{eq102}) boils down to
\beq
H_{ST}\,=\,
m + m\Phi + \dfrac{1}{2m} \left(1+3\Phi\right)p^2
\,- \,\dfrac{3i}{2m}\,(\pa_k\Phi)\,p_k
\,+\, \dfrac{3}{4m}\, \si^l \ep^{lkj} (\pa_k\Phi)\,p_j\, ,
\label{eq107}
\eeq
that has partially different numerical coefficients compared to
our result (\ref{eq101}).
We re-derived the result of \cite{Silenko-2005}, the details and
output of these calculations, together with the discussion of the
mentioned differences are reported below in subsection \ref{sec61}.

\subsection{Equations of motion for a massive particle with spin}
\label{62}

The next part is to obtain the equations of motion for a
nonrelativistic particle following the same procedure we used
in the general metric case and the gravitational wave background.
As a first step, we derive the equation for $x_i$ in
\eqref{eq51} in the case of Newtonian limit of the Schwarzschild
metric,
\beq
\dfrac{dx_{i}}{dt}
\,=\, \dfrac{1}{m} \,\Big( 1 + \dfrac{7\Phi}{2 c^2} \Big) \,\Pi_i.
\label{eq108}
\eeq
Thus, the first relativistic correction just makes a small
rescaling of the mass of the particle.

Consider the equations of motion for the momentum components
$p_i$. Let us note that this part may have certain importance related
to the experiments on the high-precision measurement of the small
ferromagnetic particles caused by the weak gravitational field
\cite{South} created by other particles.
Regardless those are macroscopic particles, the model of a massive
Dirac particle with spin may be used as a simplest approximation.
It is worth noting that the motion of small macroscopic particles
with magnetic moment was not explored theoretically, until
present\footnote{Authors are
grateful to Yu.N. Obukhov for the discussion of this point.}.

The equation (\ref{eq58}) with the metric \eqref{eq98}, becomes
\beq
\dfrac{dp_i}{dt}
\,\,=\,\,  eE_i - m(\pa_i\Phi)
\,-\, \dfrac{7}{4mc^2}\, (\pa_i\Phi)\,\vec{\Pi}^2
\,+\,
\dfrac{e}{mc}\Big(1+\dfrac{7\Phi}{2 c^2}\Big)\,(\pa_i A_l)\Pi_l .
\label{eq109}
\eeq
Once again, to make a comparison with the result of \cite{Silenko-2005},
one has to switch off the electromagnetic field and use the natural
units, that transforms (\ref{eq109}) into
\beq
\dfrac{dp_i}{dt}
\,=\,-\,m(\pa_i\Phi) \,-\, \dfrac{7}{4m}(\pa_i\Phi)\,p^2.
\label{eq110}
\eeq

The analogous equation of \cite{Silenko-2005} is
\beq
\dfrac{d\vec{p}}{dt} \, = \, \dfrac{\ep^2+p^2}{\ep}\,\vec{g},
\label{eq111}
\eeq
where $\ep$ an $\vec{g}$ are defined in \eqref{eq103a}.
Using the $\mathcal{O}(1/m)$ approximation, we obtain the equation
\beq
\dfrac{dp_i}{dt} \,=\, -\, m\,(\pa_i\Phi) \,-\, \dfrac2m\, (\pa_i\Phi)\,p^2.
\label{eq112}
\eeq
Once again, there is a numerical difference with \eqref{eq110}, which should be expected from the general situation with the
field equations, as discussed above and, also, in the literature, e.g.,
in \cite{Obukhov-2000,Silenko-2005}.
Independent on this difference, we can see from both \eqref{eq110}
and \eqref{eq112}, that an additional acceleration depends on the
orientation and magnitude of the velocity of the particle, but not
on the spin variable. This means we do not have too much
expectation to meet spin-gravity interactions for small magnetic
particles too. Anyway, it would be an interesting issue to check
this result for a more realistic model of a macroscopic particle.
The positive output could give, in principle, a qualitatively new
test of GR. Unfortunately, the result looks negative because the
acceleration of the particles does not depend on the orientation of
their spins.

The last equations describe the dynamics of the spin components
related to $\si_i$. For the metric \eqref{eq98}, the
equation for the spin components in \eqref{eq67} becomes
\beq
&&
\dfrac{d\si_{i}}{dt} \,=\,
-\,\dfrac{e}{mc} \Big(1 - \dfrac{\Phi}{2c^2}\Big)
\,\ep_{ijk}\, B_j \si_k
\,+\, \dfrac{4e}{mc^3}\,\Phi\,\ep_{ijk}B_j\si_k
\nn
\\
&&
\qquad \qquad 
+\, \dfrac{3}{2mc}\, \big[ \left( \pa_{k} \Phi \right)
\Pi_{i} \,\si_k \, + \, \left( \pa_{i} \Phi \right) \Pi_{k} \,\si_{k} \big].
\label{eq113}
\eeq
In case of zero electromagnetic field, using the natural system
of units, we arrive at
\beq
\dfrac{d\si_{i}}{dt}
\,\,=\,\, \dfrac{3}{2m}\,
\Big[\left( \pa_{k} \Phi \right) p_i \, \si_k
- \left(\pa_{i} \Phi \right) p_k\, \si_k \Big].
\label{eq114}
\eeq
The equation from \cite{Silenko-2005} (in our notations) has the form
\beq
\dfrac{d\vec{\si}}{dt}
\,=\,
\dfrac{2\ep+m}{\ep(\ep+m)}\,
\vec{\si}\times(\vec{g}\times\vec{p}).
\label{eq115}
\eeq
Performing the $\mathcal{O}(1/m)$ expansion, in the first
nontrivial order we obtain
\beq
\dfrac{d\xi_i}{dt}
\,=\,
\, \dfrac{3}{2m}\Big[
(\pa_k\Phi) \,p_i\,\si_k \,-\, (\pa_i\Phi)\, p_k\,\si_k \Big].
\label{eq116}
\eeq
It is easy to see that there is a perfect agreement with our result
(\ref{eq114}) in this case.


\subsection{On the differences between  distinct approaches}
\label{sec61}

In the previous section, we observed that our results regarding the
Newtonian limit of Schwarzschild's metric and the ones of
\cite{Silenko-2005}, have similar overall structure, but there
are discrepancies in the coefficients. This section aims clarifying
these differences. To do so, we insert the metric of the Newtonian
limit \eqref{eq98} into the general relativistic equation~\eqref{eq10}.
After that, we derive the nonrelativistic limit by using three different
methods, namely \textit{i)} the same as used in Sec.~\ref{s3} for
the general metric, \textit{ii)} the conventional perturbative
Foldy-Wouthuysen transformation (FW) \cite{FW},
as used in~\cite{Silenko-2005},
and \textit{iii)} exact Foldy-Wouthuysen transformation (EFW)
\cite{Eriksen-Kolsrud} used
in~\cite{Obukhov-2000}. Let us present these calculations in detail.

With the metric corresponding to the Newtonian limit without
electromagnetic field, and using the units with $c=\hbar=1$,
Eq.~\eqref{eq10} produces the Hamiltonian
\beq
\label{ham1}
H_{NL}
\,=\,\be m(1+\Phi)+\al^k\pi_k\,=\,\be m
+ \mathcal{E} + O,
\eeq
where we used the notation
\beq
\label{pi}
\pi_k \,=\, (1+2\Phi) \,p_k \,+\, \dfrac{i}{2}(\pa_k\Phi).
\eeq
Even
and odd
operators are
\beq
\label{operators}
\mathcal{E}=\be m\Phi
\qquad \textrm{and} \qquad
O\,=\,\al^k\pi_k.
\label{operators-newt}
\eeq
Deriving the gravitational version of Pauli equation in the same way
as in Sec.~\ref{s3}, we arrive at the result \eqref{eq101}, as expected.
In this way, we got a partial verification of the result by using an
inverted procedure of calculation.
The next step is to perform the FW transformation up to the order
$\mathcal{O}(1/m)$, using the reduced formula (see,
e.g.,~\cite{BjorkenDrell-1})
\beq
\label{FW1}
H_{FW}\,=\,
\mathcal{E} +
\be \Big(m + \dfrac{O^2}{2m}\Big)
- \dfrac{1}{8m^2}\big[O,\big[O,\vp\big]\big],
\eeq
which gives
\beq
\label{FW4}
H_{FW}
\,=\,
m + m\Phi + \dfrac{1}{2m}\big(1+3\Phi\big)\,p^2
+ \dfrac{3}{4m}\,\si^l\,\ep^{ljk}\,\big(\pa_j\Phi\big)\,p_k.
\eeq
Using the transformation of the Hamiltonian from
Ref.~\cite{Obukhov-2000},
\beq
\psi' \,=\,\Big(1-\dfrac{3}{2}\Phi\Big)\psi,
\qquad
H_{ST}\,=\,\Big(1-\dfrac{3}{2}\Phi\Big)\,H_{FW}
\,\Big(1+\dfrac{3}{2}\Phi\Big)\,,
\eeq
the equation \eqref{FW4} meets perfect agreement with
\eqref{eq107}, while being still different from the result
for the  Pauli-like equation \eqref{eq101}.

As a last verification, we perform the EFW transformation.
This method demands that, for the operator $J\equiv i\ga^5 \be$,
the Hamiltonian satisfies $\{H,J\}=HJ+JH=0$.  This is true for
Eq.~\eqref{ham1}.
Making the transformation of Hamiltonian by the formula
\beq
H_{EFW}=\{\sqrt{H^2}\}\,\,\be\,+\,\{\sqrt{H^2}\}\,\,J\,,
\eeq
we arrive at the result
\beq
\label{eq.EFW}
&&
H_{EFW}
\,=\, m +  m\Phi + \dfrac{1}{2m} \big(1+3\Phi\big)\,p^2
- \dfrac{i}{2m} \big(\pa_k \Phi\big)\, p_k
\nn
\\
&&
\qquad \qquad
+ \,\dfrac{1}{m} \ep^{kjl} \big(\pa_k \Phi\big)\, p_j  \si_l
- \dfrac{\si_k}{2}\,\big(\pa_k \Phi\big).
\eeq
It is easy to see that all the terms in this expression have the same
structure, but different coefficients, compared to the results from
the previous methods, except for the last term which is a
qualitatively new one.
This last term is the spin contribution found on~\cite{Peres}, which,
according to \cite{Obukhov-2000}, is only expected to appear as a
consequence of the application of this method.

With all this, the derivations performed by different authors
are confirmed in all cases, when the same methods were used,
in all three cases. This means we cannot count on the calculational
errors in explaining the differences between the coefficients derived
by distinct methods. Thus, it makes sense to analyse the consistency
of the utilized approximations in all three cases. The described
discrepancies in the coefficients may signal that there is a kind of
a real problem, as discussed in \cite{Obukhov-2000}.
For the perturbative FW method, the idea is to remove
the odd part of the Hamiltonian operator by making unitary
transformations. The solution appears in the form of perturbative
power series in the parameter representing the ratios of the
interaction terms, between the spinor and external field, and the
mass of the particle.  However, in the case of a gravitational
external field and in the Newtonian limit, there is a $m\Phi$
term in the $\mathcal{E}$ operator in \eqref{operators}.
This means, the  interaction term is proportional to the mass
of the particle.  Similar situations take place for the Pauli
equation and the EFW transformation. At the moment when
we take the non-relativistic limit, the power series of our interest
correspond to the parameter proportional to $m/m =1$ and the
first-order in interaction approximation goes out of the control.
Thus, the weak relativistic corrections in all three cases are not
uniquely defined and this explains the difference in the
coefficients (let us note that a possible convergence problem in the
$\mathcal{O}(1/m)$ approximation for the SFW transformation
was previously noted in \cite{Obukhov-2000}). Another source
of ambiguity is that the unitary transformations in the SFW and
EFW cases may be different, that results in the different results
of distinct procedures of arriving to the Newtonian limit. This
aspect of the theory has been discussed in
\cite{Silenko-2005,Silenko-2006}\footnote{We are grateful to
A. J. Silenko and O. V. Teryaev for mentioning this important
detail of \cite{Silenko-2005}.}.

Another possible origin of the ambiguity is related to the
non-Hermiticity of the Dirac Hamiltonian in curved spacetime
\cite{HuangParker}. This subject was extensively discussed,
e.g., in Refs.~\cite{Armi,Gorbatenko-2010}, see also the recent
review \cite{Vergeles-2022} (this review contains many other
relevant considerations). It might happen that the unitary
transformation to the diagonal basis, i.e., separating big and
small components of the Dirac spinor, is not unique because of
the non-Hermiticity of the initial operator. Our general impression
is that the general solution to the problem of Hermiticity is unknown
at this moment.

Summarizing our comparisons, at the present state of knowledge
there is a real ambiguity in
the numerical coefficients (but, in almost all cases, not in the
functional dependencies which we called structures), which are
typical for the limit of Newtonian metric. The ambiguity extends
to all three available methods of deriving a nonrelativistic limit
for the Newtonian background.
The aforementioned ambiguity concerns the metrics when the
deviation from Minkowski space is proportional to the rest mass
of the particle. This includes the Newtonian background and its
extensions to the motion in non-inertial frames, such as the
case of rotation. The last case we did not elaborate in the present
work. In many other cases, the result
is unique. E.g., using the metric of a weak plane gravitational
wave propagating along the axis $OX$ \eqref{eq68} into the
general relativistic equation \eqref{eq10}, we arrive at the
Hamiltonian
\beq
H_{GW}=\be m+ e\phi+\big(\al^k+\up\,T_{kj}\,\al^j \big)\,\Pi_k,
\label{GWagain}
\eeq
where $T_{jk} \,=\, \diag(0, -1, 1)$ and  $\up$ is a function of the
light cone coordinate $ct - x$. We can identify the even and
odd operators as
\beq
\mathcal{E}\,=\, e\phi
\qquad \textrm{and} \qquad
O \,=\, \big(\al^k+\up\,T_{kj}\al^j \big)\Pi_k.
\label{operators-gw}
\eeq
One can see that for this kind of the gravitational contribution,
the even operator $\mathcal{E}$ does not carry mass dependence.
This explains why the results coming from distinct methods
are in a perfect agreement, as we saw in Sec.~\ref{s5}.

\section{Conclusions}
\label{ConcDisc}

The nonrelativistic approximation for the Dirac equation was
considered for a combination of a weak gravitational field with 
a gauge fixing condition related to the
synchronous gauge and a general weak electromagnetic field.
The calculations are technically more complex and cumbersome
than the ones for particular metrics, such as the plane gravitational
wave or the Newtonian gravity approximation.  Our main
calculations were performed using a more simple and ``basic''
method, compared to the perturbative or exact  Foldy-Wouthuysen
transformations. The main result are the extended version of the
Pauli equation (\ref{eq38}) and Eqs.~(\ref{eq51}), (\ref{eq61})
and (\ref{eq67}) for a charged spinning particle.

In the particular case of the plane gravitational wave we met a
perfect correspondence with the previous equations of
\cite{Quach-2015}, which partially corrected the first derivation
of \cite{Goncalves-2007}.  In another particular case, of the
Newtonian gravity, we found the same general structure of the
nonrelativistic limit that was previously reported in
\cite{Obukhov-2000,Silenko-2005,Obukhov-2009} and other
publications, but in most cases, with different coefficients.
We have shown that this discrepancy is not a result of calculational
errors, but a manifestation of the ambiguity coming from the
nature of the approximation scheme. All known forms of the
nonrelativistic approximations are based on the expansions in
powers of interaction energy divided by the mass (in $c=1$ units).
Since in the Newtonian approximation the gravitational potential
is also proportional to the mass of the test particle, the
ambiguity emerges.
This explanation is confirmed by the mentioned fact that, for the
plane gravitational wave, there is no ambiguity problem and the
results show an agreement with the previous derivations which
used two types of the Foldy-Wouthuysen transformations
\cite{Quach-2015}. Another possible source of ambiguity is the
non-Hermiticity of the curved-space Dirac Hamiltonian
\cite{HuangParker}. As a result, for Newtonian regime or (as
expected) in an arbitrary non-inertial frame, we have a qualitatively
safe structure of equations, but without the uniquely-defined
coefficients.
From the qualitative side, an important result is that the relativistic
corrections to the Newton's law depend on the velocity of the
particles, but does not depend on the spin of the particles. Thus,
the situation is qualitatively  similar to the one of MOND
\cite{MOND}, which is tested in the currently running experiments
\cite{South}.
Finally, we can note that
in the recent publications \cite{Khriplovich-1997,Oancea}, the
semiclassical equations
for a spinning particle was obtained using WKB method without
nonrelativistic approximation. It would be certainly interesting to
explore the nonrelativistic approximation of these equations,
including the possible ambiguities, and compare with the results
of the present and other works on the subject.

\section*{Acknowledgements}

Authors are grateful to A. J. Silenko, O. V. Teryaev and especially to
Yu. N. Obukhov for useful discussion. S.W.P.O.
is grateful to Coordena\c{c}\~{a}o de Aperfei\c{c}oamento de Pessoal
de N\'{i}vel Superior—CAPES (Brazil) for supporting his Master's
Degree project and his current Ph.D. project. G.Y.O.  is grateful to
Coordena\c{c}\~{a}o de Aperfei\c{c}oamento de Pessoal de N\'{i}vel
Superior—CAPES (Brazil) for supporting this Master's Degree project
and to Funda\c{c}\~{a}o de Amparo \`a Pesquisa do Estado do
Esp\'{i}rito Santo -- FAPES (Brazil) for supporting his current Ph.D.
project. The work of I. Sh. is partially supported by Conselho Nacional de
Desenvolvimento Científico e Tecnológico - CNPq under the grant
305122/2023-1.

\vskip 4mm



\end{document}